# Discovery of transient topological crystalline order in optically driven SnSe

Masataka Mogi[1,2,*], Dongsung Choi[3,4,*], Kyoung Hun Oh[1], Diana Golovanova[5], Yufei Zhao[5], Yifan Su[1], Zongqi Shen[1], Doron Azoury[1], Haoyu Xia[1], Batyr Ilyas[1], Tianchuang Luo[1], Noriaki Kida[6,7], Taito Osaka[6,7], Tadashi Togashi[6,7], Binghai Yan[5] & Nuh Gedik[1✉]

[1]Department of Physics, Massachusetts Institute of Technology, Cambridge, MA, USA.

[2]Department of Applied Physics, University of Tokyo, Bunkyo-ku, Tokyo, Japan.

[3]Department of Electrical Engineering and Computer Science, Massachusetts Institute of Technology, Cambridge, MA, USA.

[4]Max Planck Institute for the Structure and Dynamics of Matter, Hamburg, Germany.

[5]Department of Condensed Matter Physics, Weizmann Institute of Science, Rehovot, Israel.

[6]RIKEN SPring-8 Center, Sayo, Hyogo, Japan.

[7]Japan Synchrotron Radiation Research Institute, Sayo, Hyogo, Japan.

*These authors contributed equally: Masataka Mogi, Dongsung Choi.

✉e-mail: gedik@mit.edu

**Abstract**

**Ultrafast optical excitation provides a powerful route for accessing emergent quantum phases far from equilibrium, enabling transient light-induced phenomena such as magnetism, ferroelectricity, and superconductivity. However, extending this approach to induce topological phases, especially in conventional semiconductors, remains challenging. Here, we report the observation of a thermally inaccessible, transient topological crystalline order in the layered semiconductor SnSe, a trivial insulator with a sizable (~ 0.8 eV) band gap, induced by femtosecond above-gap excitation. Time- and angle-resolved photoemission spectroscopy directly reveals the sub-picosecond emergence of Dirac-like linear dispersions within the band gap. Their features, including a high Fermi velocity (~ $2.5\times10^5$ m/s), multiple Dirac points away from high-symmetry momenta, and independence from probe photon energy, are consistent with mirror-symmetry-protected surface states of a topological crystalline insulator. The observed spectral dynamics, combined with density functional theory calculations, indicate that the femtosecond excitation transiently increases lattice symmetry, enabling topological crystalline order to emerge. Our discovery opens new avenues for ultrafast optical control of topological quantum phases in semiconductors, with potential applications in quantum and spintronic devices.**

## Main

The discovery of quantum phases governed by global electronic topology has fundamentally reshaped our understanding of condensed matter systems, revealing robust edge and surface states protected by symmetries such as time-reversal or crystalline symmetry[1,2]. Experimental realizations, including topological insulators and Dirac/Weyl semimetals, have so far centered on narrow-gap semiconductors or semimetals[1–8], where strong spin-orbit coupling drives band inversion. By contrast, large-gap semiconductors tend to remain topologically trivial in equilibrium, as their electronic structures are more rigid and less responsive to external fields. However, this rigidity could be advantageous for device applications: a fully insulating bulk cleanly distinguishes topological transport channels, provided that a viable route to induce topology can be found.

A promising strategy to overcome this rigidity is to drive the system far from equilibrium[9–12]. In particular, femtosecond laser light has emerged as a powerful tool to reshape lattice symmetry and electronic structure on sub-picosecond timescales (Fig. 1a,b), thereby opening non-thermal pathways to emergent states[13,14]. Indeed, ultrafast light-matter interaction has been shown to induce quantum phenomena such as light-induced superconductivity[15] and ferroelectricity[16,17] that are absent in equilibrium. In the context of topology, ultrafast pulses have enabled modulation or suppression of existing topological states[18–23], for instance, via Floquet engineering in topological surface states[18]. However, a critical open question remains: can one create a nontrivial topological phase "from scratch" in a system that is trivial under equilibrium conditions?

To address this question, we focus on SnSe, a layered IV-VI semiconductor with a sizable band gap of ~0.8 eV[24]. While several rocksalt IV-VI compounds, such as SnTe and (Pb,Sn)Se, are well-known TCIs protected by mirror symmetry[4,25,26], SnSe is topologically trivial under equilibrium conditions. At ambient temperature, SnSe crystallizes in an orthorhombic Pnma phase (Fig. 1c), where puckered Sn-Se bilayer stacking, stabilized by stereochemically active Sn 5s lone-pair electrons[27], breaks mirror symmetry, resulting in a topologically trivial gapped state. At elevated temperatures (~800 K), SnSe undergoes a transition to a higher-symmetry Cmcm phase[28–30] (Fig. 1d), known for exceptional thermoelectric properties[24,27]. However, this phase also remains topologically trivial, since the lone pair orbitals persist and continue to break the symmetry crucial for band inversion.

Nevertheless, recent studies suggest that nonequilibrium methods can unlock topological phases in SnSe. For instance, epitaxial growth on $Bi_2Se_3(111)$[10,11,31] has been shown to stabilize SnSe in a cubic rocksalt structure, analogous to the prototypical TCI ($Pb_{1-x}Sn_x$)Se ($x \leq 0.4$)[4,25,26]. Furthermore, ultrafast x-ray diffraction experiments have indicated that femtosecond laser excitation may transiently drive SnSe toward a rocksalt-like Immm phase[32] (Fig. 1e), characterized by shear-like atomic displacements opposite to those in the high-temperature Cmcm phase transition. While previous results provide valuable insight into the structural dynamics of photoexcited SnSe[32–35], they leave open the question of whether the transient phase hosts nontrivial topological properties. Recent theoretical works[36,37] propose that relaxation from the Immm phase can lead to a stabilized rocksalt structure that hosts TCI order, but the topological nature of the transient phase itself remains to be established[32–35,38,39].

Here, by combining density functional theory (DFT) calculations and femtosecond time- and angle-photoemission spectroscopy (trARPES) with near-infrared laser excitation (100 fs, 1.55 eV), we provide direct evidence that ultrafast optical excitation induces Dirac-like surface states supported by nonzero mirror Chern numbers. This transient state emerges within the band gap on subpicosecond timescales - even at room temperature - demonstrating a robust topological crystalline order not achievable under equilibrium conditions.

**Lattice and electronic structures**

To elucidate the electronic properties of the transient Immm phase, we performed DFT calculations using the orthorhombic lattice parameters at ambient conditions ($a$ = 11.5 Å, $b$ = 4.19 Å, $c$ = 4.39 Å; Supplementary Section 1), under the assumption that the lattice size does not vary within a few picoseconds after excitation. The calculated band structures for the (100) surface planes of the Pnma, Cmcm, and Immm phases are shown in Fig. 1f-h. In the Pnma phase (Fig. 1f), the lack of mirror symmetry results in a sizable band gap (~0.8 eV), consistent with its trivial insulating nature. The high-temperature Cmcm phase (Fig. 1g), despite its higher symmetry, remains a gapped (~0.4 eV), topologically trivial state. In stark contrast, the hypothetical Immm phase (Fig. 1h) exhibits substantially altered band structures with multiple gapless crossings near the $\overline{\mathrm{Y}}$ and $\overline{\mathrm{Z}}$ points.

In this Immm phase, our calculations identify multiple gapless Dirac cones located along $\overline{\Gamma}-\overline{\mathrm{Z}}$ (near $\overline{\mathrm{Z}}$) and $\overline{\Gamma}-\overline{\mathrm{Y}}$ (near $\overline{\mathrm{Y}}$) on the (100) surface, each exhibiting linear dispersion with Fermi velocities of ~$2.5\times10^5$ m/s. Unlike time-reversal-invariant topological insulators[1,2,40], where Dirac cones are typically found at high-symmetry points such as $\overline{\Gamma}$ or $\overline{\mathrm{X}}$, these cones robustly emerge away from such points, resembling the behavior of other TCI materials in the (Pb,Sn)Se family[4,25,26]. Crucially, these Dirac cones appear within the band gap of SnSe, in stark contrast to the fully gapped Pnma and Cmcm phases, and remain robustly gapless even with slight residual distortions (Supplementary Sections 11 and 12). This unique gapless surface states inside the bulk gap makes them readily detectable by trARPES.

**Ultrafast observation of transient in-gap band dispersion**

We performed trARPES measurements on 50-nm-thick SnSe(100) films epitaxially grown on InP(100) (sample #1, see Methods). Using 10.8 eV extreme ultraviolet (XUV) probe pulses[41], we accessed the electronic states around the $\overline{\mathrm{Y}}$ and $\overline{\mathrm{Z}}$ points in the two-dimensional Brillouin zone[38,42] (Fig. 2b,c). These momentum regions were chosen based on DFT predictions indicating that Dirac surface states would emerge in the transient Immm phase (Fig. 1h).

Figure 2a presents energy-momentum snapshots measured along the $\overline{\Gamma}-\overline{\mathrm{Z}}$ direction at selected pump-probe delays. Before photoexcitation (pump-probe delay $\Delta t$ = –1.4 ps), the spectral weight near the Fermi level ($E_\mathrm{F}$) is nearly absent, reflecting the intrinsic insulating nature of SnSe in the equilibrium Pnma phase. Upon above-gap femtosecond photoexcitation (1.55 eV, 100 fs pulse width) at an incident fluence of 1.0 mJ/cm$^2$ (nominal absorbed fluence

~ 0.4 mJ/cm$^2$), a linearly dispersive in-gap feature with negative group velocity ($\partial E/\partial k_z < 0$) emerges within 0.2 ps, reaching maximum intensity around 0.6 ps, and subsequently decaying over ~ 5 ps. This highly dispersive feature is incompatible with nondispersive defect states or localized mid-gap states. Moreover, the observed signal lies entirely below $E_F$, ruling out the direct occupation of photoexcited conduction band states[38], as such states would appear above $E_F$ in equilibrium. The picoseconds lifetime of the transient state is distinct from Floquet-Bloch states or laser-assisted photoemission[18,38], which occur only during the pump-probe temporal overlap, as well as from extrinsic effects such as surface photovoltage[43], which typically persist on hundreds of picoseconds timescales. Furthermore, our DFT calculations confirm the absence of in-gap states in both the Pnma (Fig. 1f) and Cmcm (Fig. 1g) equilibrium phases, reinforcing that the observed feature originates from a genuine, non-thermal emergent phase.

Reproducibility of this transient feature was confirmed across multiple SnSe films, and its persistence up to room temperature was also observed (Supplementary Section 8). Further exclusion of alternative interpretations is discussed in Supplementary Section 13.

**Evidence for transient topological crystalline order**

To verify the topological nature of the emergent in-gap band, we systematically examined its momentum-space location and symmetry properties. First, the transient linear dispersion identified along the $\bar{\Gamma} - \bar{\mathrm{Z}}$ direction (Fig. 2a) is also observed along the perpendicular $\bar{\Gamma} - \bar{\mathrm{Y}}$ direction (Fig. 2d), indicating that multiple linearly dispersive states coexist within the Brillouin zone. This observation aligns with our DFT calculations, which predict the presence of Dirac cones near the $\bar{\mathrm{Y}}$ and $\bar{\mathrm{Z}}$ points (Fig. 1h). The extracted Fermi velocity (~$2.5\times10^5$ m/s) is comparable to those of established topological systems such as $Bi_2Se_3$ (~$5\times10^5$ m/s, ref. [40]) and SnTe (~$3\times10^5$ m/s, ref. [25]).

To decisively identify these dispersive features as Dirac cones, we searched for the Dirac point by performing additional trARPES measurements using a higher probe photon energy (21.6 eV), providing access to a larger momentum range. We used a slightly different SnSe sample (sample #2), grown on a PbSe(100) buffer layer. Although this buffer layer introduces a minor chemical potential shift compared to sample #1 (Fig. 2a), the sample quality remained comparable (Supplementary Section 1). These measurements revealed both positive ($\partial E/\partial k_z > 0$) and negative group-velocity branches, allowing us to identify a Dirac point at the intersection of the two branches at $k \approx 0.92$ Å$^{-1}$ (Fig. 3a).

Importantly, as shown in Fig. 3b, these Dirac-like bands remain fixed in momentum space regardless of probe photon energy, a hallmark behavior of two-dimensional surface states, in contrast to three-dimensional bulk valence bands which shift in momentum/energy position due to their out-of-plane dispersion ($k_x$)[37]. The lack of observable Dirac cones in the first Brillouin zone can be understood to result from matrix element effects, similar to those observed in related TCIs, such as our data for ($Pb_{1-x}Sn_x$)Se ($x = 0.23$) (Fig. 3b, inset) as well as SnTe (ref. [25]), where the Dirac branch in the first Brillouin zone could be significantly suppressed depending on probe photon energy. These observations suggest that SnSe likely hosts a similar double-cone structure (Fig. 3c), with the inner cone remaining unresolved under the photon energies employed here (10.8 eV and 21.6 eV).

Notably, the experimentally identified Dirac point is shifted by $\delta k \approx 0.20$ Å$^{-1}$ from the zone boundary (Fig. 3b). This displacement is considerably larger than the ~0.08 Å$^{-1}$ shift observed in (Pb,Sn)Se and is consistent with trends reported in (Pb,Sn)Te, where increasing Sn content progressively shifts the Dirac point away from the zone boundary[44] (Supplementary Section 9). Moreover, DFT calculations clarify that the bulk valence and conduction band edges are located at the zone boundary (Fig. 3d), making it unlikely that the observed transient band originates from bulk states. Taken together, these comprehensive measurements and analyses conclusively demonstrate that the photo-induced transient phase of SnSe exhibits the hallmark characteristics of a TCI.

## Discussion and outlook

Our discovery of ultrafast, photoinduced topological order in SnSe represents a major advance in accessing quantum states of matter. We observe the emergence of Dirac-like in-gap states on subpicosecond timescales, persisting at room temperature, far beyond the reach of conventional methods that usually rely on chemical doping. This result not only uncovers a new topological state in SnSe but also sheds light on the structural and electronic dynamics that drive this nonequilibrium transformation.

Previous time-resolved X-ray diffraction (trXRD) experiments at absorbed fluences of ~2.2 mJ/cm$^2$ suggested that SnSe can be driven toward a higher-symmetry phase[32,33], yet diffraction evidence for a complete transition to the Immm structure remains elusive[35]. To look for this, we performed our own trXRD measurements at fluences up to ~1.74 mJ/cm$^2$. Similar to the earlier measurements, our results show no clear signature (Supplementary Section 10), implying that any structural transition is either partial or unrealized. In contrast, transient dynamics measured by trARPES (Fig. 4), combined with DFT calculations, provides a more detailed picture of local symmetry changes; the position of Dirac bands remains fixed regardless of pump fluence or time delay (Fig. 4c-f), contrary to the systematic Dirac cone shifts predicted by DFT under uniform lattice distortions from the Pnma to the Immm phase (Supplementary Section 12). We therefore interpret the Dirac-like bands as appearing suddenly in spatially localized regions, i.e., Immm domains forming within a background of the Pnma phase. This spatially inhomogeneous switching, in which microscopic domains govern the macroscopic response, is reminiscent of subthreshold transitions reported in correlated materials like $VO_2$ (ref. [45,46]) and charge density wave systems[47], where local order emerges before the establishment of a globally ordered phase.

On a microscopic level, our results highlight the key role of photoexcited electron depletion from lone-pair orbitals localized on Sn 5*s* states hybridized with Se 4$p_x$ orbitals[27,32]. This depletion disturbs the stereochemical activity that sustains the puckered bilayer and temporarily stabilizes a high-symmetry lattice through enhanced σ-bonding among Sn and Se *p*-orbitals, leading to the formation of the rocksalt-like Immm phase[48]. The picosecond lifetime of this phase accords with the timescale of electron-hole photoinjection dynamics (Fig. 4a). As shown in the top panel of Fig. 4b, the temporal decay of the photo-induced in-gap band intensity (red region in Fig. 4a) coincides with the depletion of valence band electrons (blue region) (see Methods for more detail). This behavior is further supported by the linear scaling of the in-gap band intensity with pump fluence (bottom panel of Fig. 4b), consistent with a photo-excited hole density that scales with pump excitation fluence.

Our findings show the potential of nonequilibrium routes to generate transient topological orders in semiconductors with sizable gaps. There exist several tuning knobs that might extend the lifetime or improve the spatial uniformity of these phases in the future: increasing pump fluence beyond current damage threshold limitations[39], engineering substrate strain[10,11], or introducing Te/Pb dopants that foster higher-symmetry structures[26,44]. More broadly, this work demonstrates that photoinduced symmetry restoration in layered semiconductors opens new opportunities for on-demand topological control at room temperature. We speculate that similar approaches could be adapted to a variety of materials [48], thus broadening the scope of quantum device engineering.

**Acknowledgments**

We are grateful to Yijing Huang, Kenji Yasuda, and Ilya Belopolski for insightful discussions. We are also thankful to Bryan Fichera and Honglie Ning for discussions about trXRD data analysis. This work was supported by the US Department of Energy, BES DMSE (data taking, analysis and manuscript writing), and Gordon and Betty Moore Foundation's EPiQS Initiative Grant No. GBMF9459 (instrumentation). M.M. is supported by JST PRESTO Grant No. JPMJPR23HA and JSPS KAKENHI Grant No. JP24K16986. N.K. is supported by JSPS KAKENHI Grant No. JP23K26157. The trXRD experiments were performed at the BL3 of SACLA with the approval of the Japan Synchrotron Radiation Research Institute (JASRI) Proposal No. 2024A8006.

**Author contributions**

N.G. supervised the project. M.M., D.C. and K.H.O. conceived the study. M.M. and D.C. grew and characterized the samples with help of H.X. M.M. and D.C. performed trARPES measurements and analyzed the data with help of H.X., D.A., and Y.S. D.G., Y.Z. and B.Y. performed DFT calculations, analyzed and interpret the data. D.C., K.H.O., M.M., Z.S., and Y.S. conducted trXRD measurements and analyzed the data with help of N.K., T.O. and T.T. B.I. and T.L. performed time-resolved reflectivity measurements and analyzed the data. M.M., D.C. and N.G. interpreted the results and wrote the manuscript with inputs from B.Y. and all other authors.

**Competing interests**

The authors declare no competing interests.

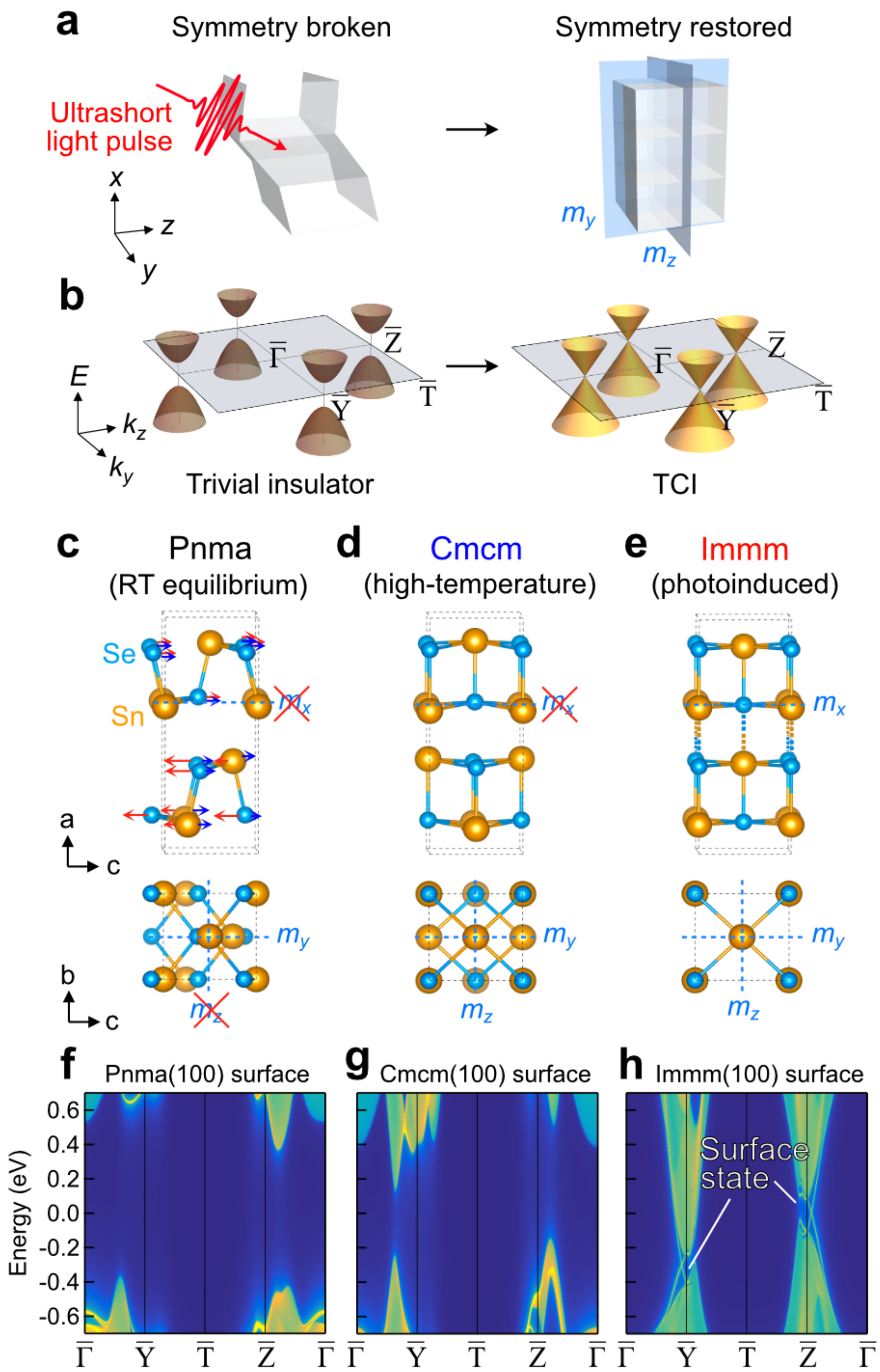

**Fig. 1 | Crystal structure, band topology, and pathway for photo-induced topological crystalline order in SnSe. a**, Schematic illustration of a photo-induced structural phase transition from a lattice-distorted state to a symmetry-restored state with mirror planes of $m_y$ and $m_z$. **b**, Band structure illustration showing that lattice distortion opens a gap, resulting in a topologically trivial insulating state, while photoexcitation restoring symmetry enables topological crystalline order. **c-e**, Crystal structures of SnSe viewed along *b*-axis (top) and *a*-axis (bottom) for orthorhombic space groups Pnma (**c**), Cmcm (**d**), and Immm (**e**). The lattice

parameters are fixed at $a$ = 11.5 Å, $b$ = 4.19 Å, and $c$ = 4.39 Å. The Pnma phase lacks a mirror plane ($m_z$), whereas the Cmcm and Immm phases possess mirror planes $m_y$ and $m_z$. Arrows indicate the atomic displacements associated with transitions to the Cmcm (blue) and Immm (red) phases. **f-h**, Calculated band structures at the (100) surface for the Pnma (**f**), Cmcm (**g**), and Immm (**h**) phases. Unlike the Pnma and Cmcm phases, the Immm phase hosts gapless surface states characteristics of a TCI.

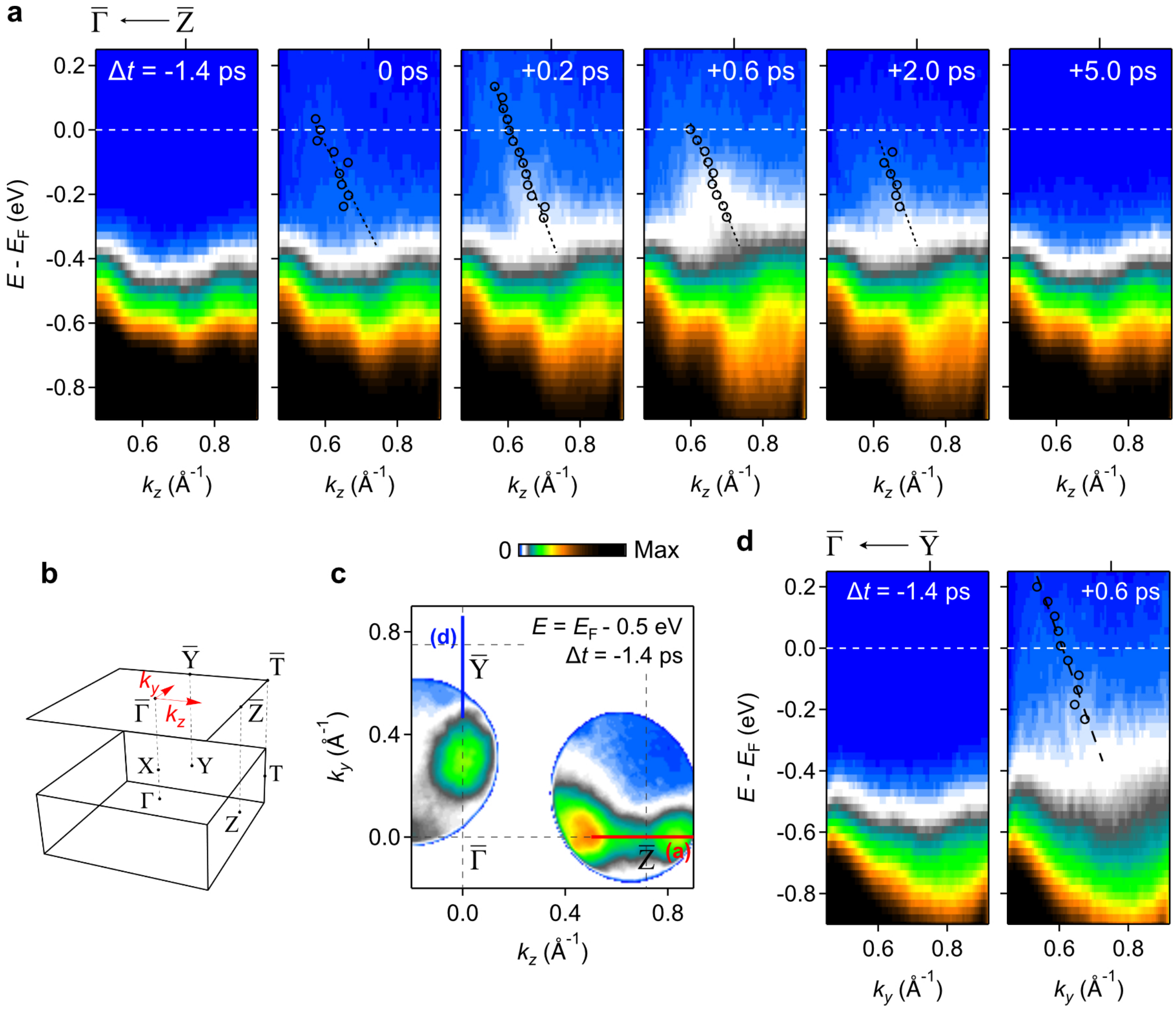

**Fig. 2 | Emergence of a transient in-gap band dispersion. a**, trARPES energy-momentum snapshots along $k_z$ at $k_y = 0$ Å$^{-1}$ near the $\bar{Z}$ point on the (100) surface of a SnSe film (sample #1), taken at various pump-probe time delays. Measurements were performed using a 10.8 eV probe pulse and a near-infrared (1.55 eV) pump pulse at an incident fluence of 1.0 mJ/cm$^2$. **b**, Bulk and projected surface Brillouin zones of the orthorhombic lattice, with key symmetry points and axes relevant to the measurements. **c**, Constant-energy cut at $E = E_F - 0.5$ eV around the $\bar{Y}$ and $\bar{Z}$ points, acquired before photoexcitation ($\Delta t$ = -1.4 ps). Dashed lines denote the boundaries of the surface Brillouin zone. **d**, trARPES maps around the $\bar{Y}$ point along $k_y$ at $k_z$ = 0.72 Å$^{-1}$, measured at $\Delta t$ = -1.4 ps (left) and $\Delta t$ = 0.6 ps (right). In a and d, the transient in-gap band is marked by black circles and dashed lines, indicating the maxima of the momentum distribution curves and the extracted linear dispersion. All measurements were performed at 45 K.

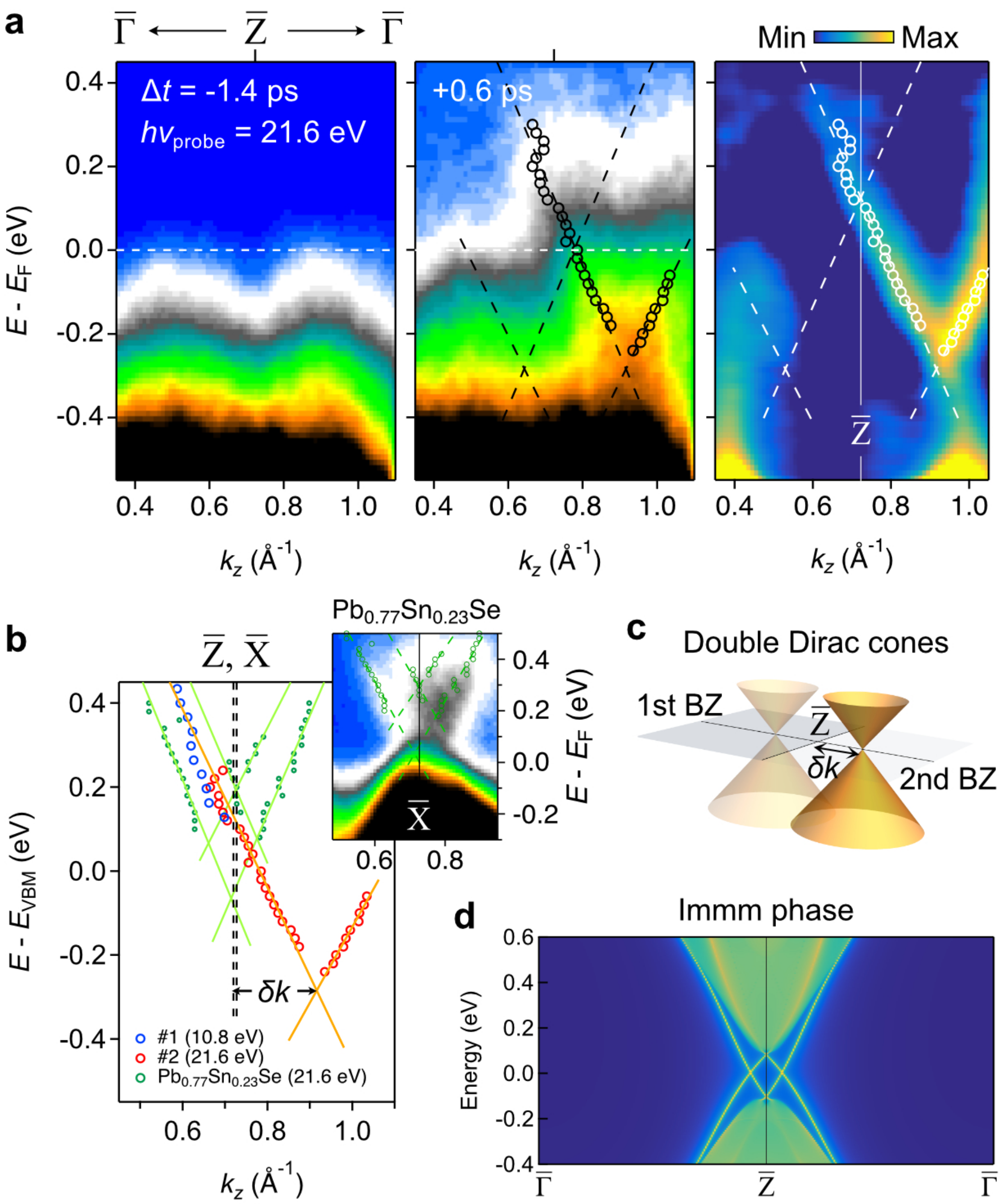

**Fig. 3 | Evidence of topological crystalline order. a**, trARPES maps around the $\bar{Z}$ point on the (100) surface (sample #2), measured using a probe photon energy of $h\nu_{probe}$ = 21.6 eV. The left and middle panels correspond to $\Delta t$ = -1.4 ps and $\Delta t$ = 0.6 ps, respectively, while the right panel shows the curvature-enhanced map at $\Delta t$ = 0.6 ps. Measurements were performed at 296 K with an incident pump fluence of 1.0 mJ/cm$^2$. The transient Dirac-like band is indicated by black circles and dashed lines, representing the extracted band positions and their linear fit, respectively. The fit is symmetrized about the $\bar{Z}$ point. **b**, Band dispersion extracted from trARPES data. Red circles correspond to peak positions obtained from momentum distribution curve analysis at 10.8 eV, while blue circles are extracted from the curvature-enhanced[49] map at 21.6 eV. Green circles show the band dispersion for $Pb_{0.77}Sn_{0.23}Se$ in the TCI phase (inset).

Solid lines represent linear fits to the data. **c**, Schematic of the electronic structure of a TCI near the $\bar{\mathrm{Z}}$ point. Two Dirac cones appear along the $\bar{\Gamma}-\bar{\mathrm{Z}}$ direction, separated by a momentum offset δ*k*. The Dirac cone in the second Brillouin zone is highlighted to illustrate the matrix element effect. **d**, DFT-calculated surface band structure of the Immm phase of SnSe around the $\bar{\mathrm{Z}}$ point.

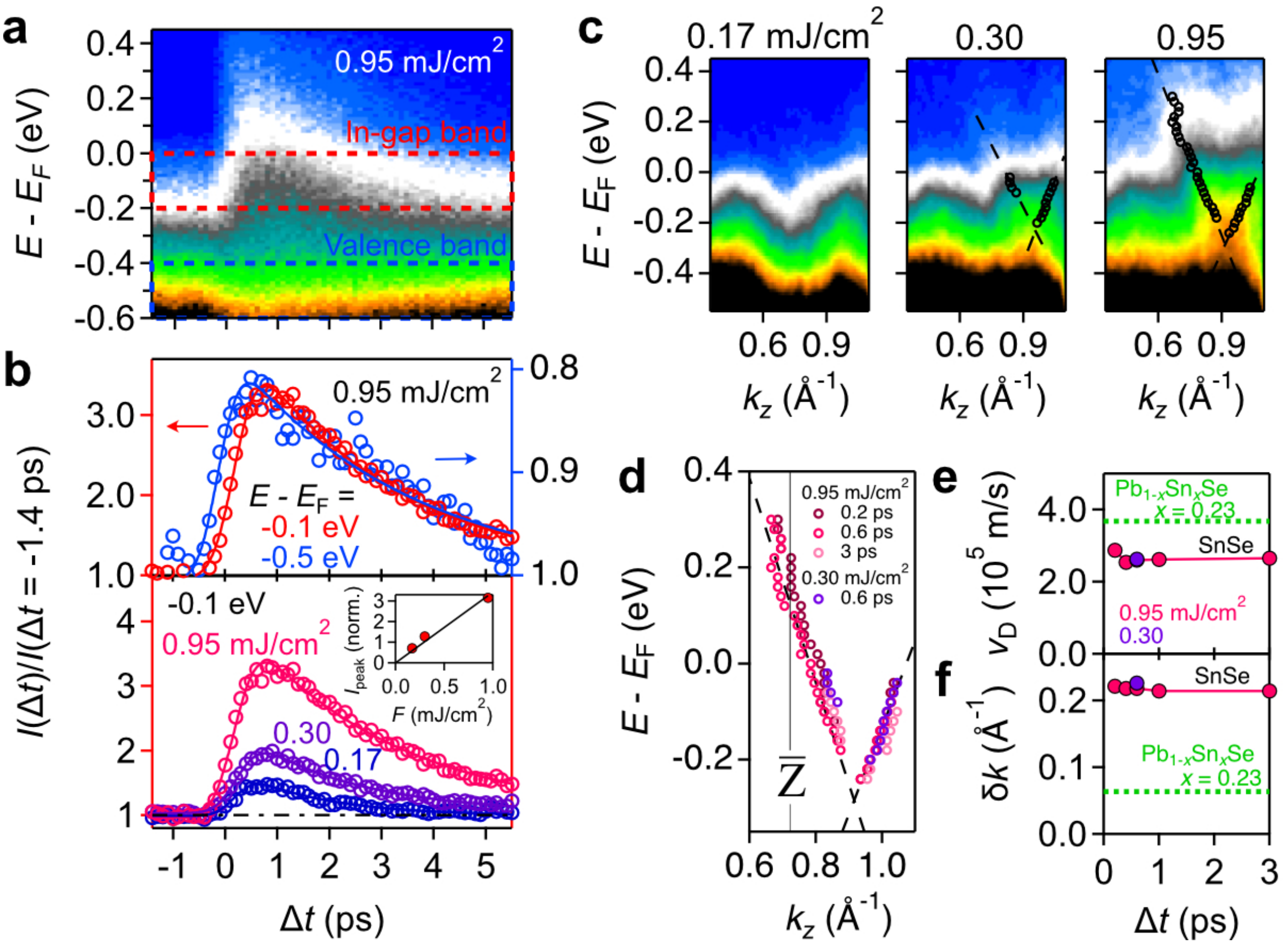

**Fig. 4 | Dynamics of transient topological crystalline order. a**, Time evolution of trARPES intensity integrated around the $\bar{Z}$ point on the (100) surface (sample #2). **b**, Time evolution of spectral weight below the Fermi level, normalized by the value at $\Delta t$ = -1.4 ps. Red (-0.1±0.1 eV) and blue (-0.5±0.1 eV) boxes indicate the energy regions of interest. The top panel compares normalized intensities for the red and blue regions at a pump fluence of 0.95 mJ/cm$^2$. The bottom panel shows the integrated intensity over the red region for selected pump fluences normalized to the value at $\Delta t$ = -1.4 ps. Solid curves represent fits to a single-exponential decay model (see Methods). The inset displays the fluence dependence of the peak intensity of the single-exponential fit ($I_{\text{peak}}$). The black line is a linear fit. **c**, trARPES band structure maps around the $\bar{Z}$ point for selected pump fluences at $\Delta t$ = +0.6 ps. **d**, Band dispersions extracted at various pump-probe time delays and pump fluences. **e**,**f**, Time evolution of the Dirac velocity ($v_D$) (e) and the momentum offset of the Dirac point relative to the $\bar{Z}$ point ($\delta k$) (f) for pump fluences of 0.30 mJ/cm$^2$ and 0.95 mJ/cm$^2$. These values are compared with those for $Pb_{0.77}Sn_{0.23}Se$ (see Fig. 3d, inset and Supplementary Section 9).

## Methods

### Sample synthesis.

The SnSe and (Pb,Sn)Se thin films were epitaxially grown on semi-insulating InP(100) substrates (>5×$10^6$ Ω·cm) in our custom-built molecular beam epitaxy (MBE) chamber (<1×$10^{-7}$ torr) at a growth temperature of 175°C. The substrates were pre-annealed at 340°C in the growth chamber prior to growth. Beam fluxes were calibrated using a quartz crystal microbalance, with an excess flux of Se, approximately twice that of Sn and Pb, to suppress Se vacancies. The film growth rate was about 0.5 nm $min^{-1}$, calibrated by X-ray reflectivity measurements. The typical thickness of our samples is about 50 nm, a thickness comparable to the optical penetration depth at our pump wavelength of 1030 nm (1.2 eV)[50]. After growth, the samples were cooled to approximately 100°C under a continued supply of Se flux, then transferred to the ARPES chamber (<1×$10^{-10}$ torr) directly connected to the MBE chamber through a preparation chamber (<5×$10^{-10}$ torr) without exposure to the atmosphere. To avoid charging the samples on the insulating substrates, they were electrically grounded by placing the tip of a tungsten wire on the sample, which was connected to the sample holder.

### trARPES.

In our trARPES setup, a Yb fiber laser (Tangerine, Amplitude) provides fundamental beam pulses with a center wavelength of 1030 nm (1.2 eV), a duration of 135 fs, a pulse energy of 250 μJ, and a repetition rate of 300 kHz. The fundamental beam is divided into two branches for the pump and probe beams. In the probe branch, to generate extreme ultraviolet (XUV) beams, the second or third harmonic is generated through β-$BaB_2O_4$ crystals and is focused on Ar gas ejected from a gas-jet nozzle. To select the desired harmonics, such as 10.8 eV and 21.6 eV, we use an off-plane mounted grating in the XUV monochromator (McPherson Inc.), which eliminates the copropagating seed beam and other harmonics[41]. Subsequently, a toroidal mirror focuses the XUV beams on the samples. The polarization of the probe beams was linear vertical polarization. In the pump branch, the fundamental beam is sent to an optical parametric amplifier (Orpheus-HP, Light Conversion) to convert the wavelength to 800 nm (1.55 eV). The pump polarization was linear horizontal polarization. The data for linear vertical polarization is shown in Supplementary Section 7.

An angle-resolved time-of-flight (ARTOF) analyzer (ARTOF 10k, ScientaOmicron) was used to collect the emitted photoelectrons. To locate the Brillouin zone edge of SnSe in the area of the photoelectron detector for our probe photon energies, we tilted the sample by approximately 42° for 10.8 eV and 22° for 21.6 eV from the flat condition. The pump and probe beams are nearly collinear, with an incident angle of about 45° at the flat condition. The incident angles of the pump and probe beams relative to the sample are approximately 3° and 23° for 10.8 eV and 21.6 eV, respectively. Our energy resolution, which combines the resolutions of our pump, probe beams, and detector, and the estimated upper bound of the time resolution for the present study are approximately 50 meV and 420 fs, respectively (Supplementary Section 3). Curvature plots[49] were used to highlight the transient dispersive bands, as presented in Fig. 3.

**DFT calculations.**

Density-functional theory (DFT) calculations were performed using the Vienna ab initio Simulation Package (VASP) with the projector-augmented wave (PAW) method[51–53]. The exchange-correlation functional was treated using the generalized gradient approximation (GGA) parametrized by Perdew-Burke-Ernzerhof (PBE)[54]. The kinetic energy cutoff for the plane-wave basis was set to 300 eV. The Brillouin zone integration was carried out using a 4×12×12 Γ-centered k-point mesh for the Pnma and Cmcm phases (a = 11.52 Å) and a 12×12×12 Γ-centered k-point mesh for the Immm phase (a = 5.76 Å). Maximally localized

Wannier functions (MLWFs) for Sn-5s, 5p, and Se-4p orbitals were constructed using the WANNIER90 package[55]. Surface and edge state calculations, based on the maximally localized Wannier functions, were performed using the WannierTools package[56].

**trXRD.**

Pump-probe hard x-ray diffraction experiments were carried out at the BL3 EH2 of the SPring-8 Angstrom Compact free electron LAser (SACLA), an x-ray-free electron laser facility in Japan. The probe consisted of 10 keV x-ray pulses with a duration of approximately 10 fs (full width at half maximum, FWHM). These pulses were focused through compound refractive lenses (CRL) to a spot size of approximately 28.5 μm × 29.3 μm in diameter (FWHM) and were incident at a grazing angle of 3.0° relative to the sample surface. The pump comprised a time-delayed 1.55 eV (797.7 nm) beam with a duration of about 30 fs (FWHM). The incidence angle of the pump beam varies as different Bragg peaks were measured, ranging from 6.9° to 8.9° with respect to the sample surface. The pump spot size was approximately 623.3 μm × 654.5 μm (FWHM). Measurements were performed using a Huber diffractometer, and x-ray scattering images were acquired with a 2D multiport charge-coupled device (MPCCD) detector. Both the 10 keV x-ray probe and the 1.55 eV infra-red (IR) pump pulses were linearly polarized (horizontal to the ground) and operated at a repetition rate of 30 Hz. All measurements started from the Bragg condition and were performed under the ambient condition.

**Dynamics of transient topological crystalline order.**

To further explore the temporal evolution of the photoinduced topological state, we conducted time-resolved measurements on the SnSe film grown on a PbSe(100) buffer layer (sample #2). Extended Data Fig. 1a displays a series of energy-momentum snapshots around the $\bar{Z}$ point, along with curvature analyses (Extended Data Fig. 1c). Consistent with the results from sample

#1 (Fig. 2a), a Dirac-like band dispersion appears as early as 0.2 ps, and can be detected at $\Delta t$ = 0 ps, likely due to temporal broadening pump and/or probe pulses, and then disappears within ~5 ps.

The evolution of the spectral intensity around the $\bar{Z}$ point is shown in Fig. 4a. After photoexcitation, intensity near $E_F$ (red square) is enhanced, indicating the formation of an in-gap band, while the valence-band intensity (blue square) is suppressed. The integrated intensities of these red and blue regions are plotted in Fig. 4b (top). Notably, the buildup of the in-gap state is slightly delayed (~0.3 ps) relative to the depletion of the valence band, whereas the recovery of the valence-band intensity and the decay of the in-gap band proceed in parallel. These temporal profiles fit well to an exponential decay model:

$$I(t) = \left[I_{\text{peak}} \exp\left(-\frac{t-t_0}{\tau}\right) \cdot \left\{1 + \text{erf}\left(\frac{2\sqrt{2}(t-t_0)}{w}\right)\right\}\right] * g(w_0, t) + C,$$

where $I_{\text{peak}}$, $t_0$, $\tau$, erf, $g$, $w$, $w_0$, and $C$ are the peak intensity, time zero, decay time, error function, Gaussian function, system response time, Gaussian temporal width, and the intensity before the arrival of the pump pulse, respectively.

This behavior is consistent with a carrier-density-driven process: electron-hole photoinjection initiates the phase transition followed by phonon-mediated structural reorganization that stabilizes the in-gap Dirac bands. Although the probed valence-band region lies around -0.5 eV, slightly shallower than the Sn 5$s$-Se 4$p_x$ hybridized lone-pair orbitals typically located near -1 eV (ref. [32]), the above-gap excitation energy (1.55 eV) is sufficient to excite electrons from deeper-lying states as well. Therefore, the observed depletion can be interpreted as partially reflecting the photoexcited hole population associated with lone-pair orbitals that drive the symmetry change.

We next examined the fluence dependence of this transient state. As shown in the bottom panel in Fig. 4b, the in-gap band intensity increases linearly with pump fluence up to ~0.95 mJ/cm$^2$, confirming that the photoexcited carrier density is a key factor in driving the topological transition.

Finally, we investigated how pump fluence affects both the Dirac dispersion and its temporal evolution. Fig. 4c demonstrates that the Dirac dispersion becomes clearer at higher pump fluences. In Fig. 4d,e we summarize the temporal dynamics of the in-gap Dirac band dispersion, showing that neither the Dirac velocity ($v_D$) nor the separation of the Dirac point from the $\bar{\mathrm{Z}}$ point ($\delta k$) changes significantly with time and fluence. The dispersion thus appears rigid under theses photoexcitation conditions. Such rigidity suggests that the photoinduced change to the Immm-like structure proceeds locally, in contrast to the uniform atomic displacements toward Immm, as studied by our DFT calculations (Supplementary Section 12).

**Data availability.** All relevant data within this paper are available from the authors.

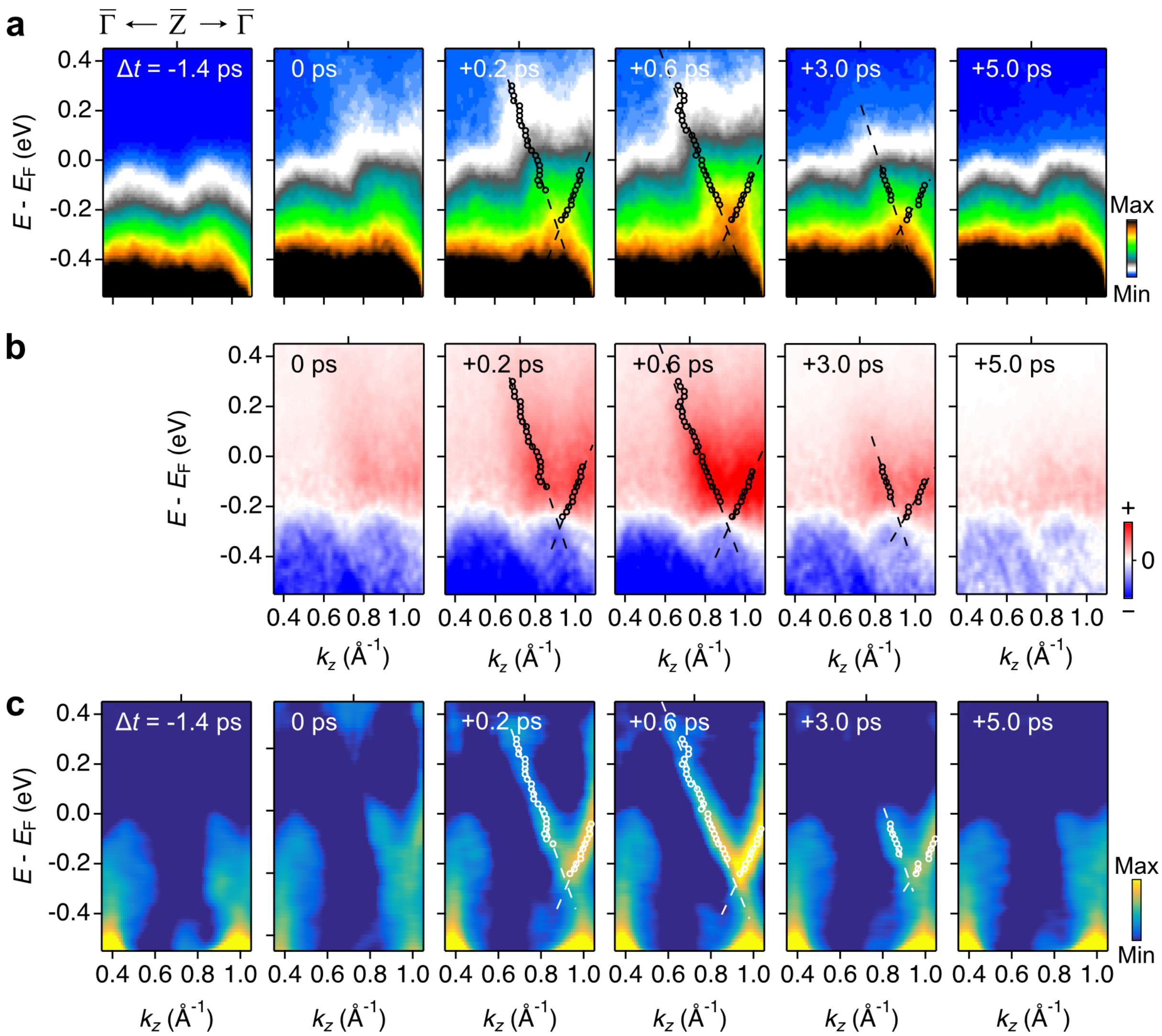

**Extended Data Fig. 1 | Transient in-gap linear dispersion for sample #2. a**, Snapshots of energy-momentum cuts along $k_z$ at $k_y = 0$ Å$^{-1}$ near the $\bar{Z}$ point at various time delays, overlaid with band dispersions (black circles) and their linear fits (dashed lines) extracted from the corresponding curvature-enhanced maps shown in (c). **b**, Difference maps of trARPES intensity relative to $\Delta t = -1.4$ ps. Red and blue indicate gain and loss of photoelectron intensity, respectively. **c**, Curvature-enhanced maps corresponding to the energy-momentum cuts in (a).